\documentclass[apj]{emulateapj}
\usepackage{amsmath}
\usepackage{natbib}
\usepackage{amsthm} 
\usepackage{amssymb}	
\usepackage{graphicx}

\newcommand{\gv}[1]{\ensuremath{\mbox{\boldmath$ #1 $}}} 
\renewcommand{\div}[1]{\gv{\nabla} \cdot #1} 

\begin{document}

\title{Quantifying the Impact of Cosmological Parameter Uncertainties on 
Strong Lensing Models With an Eye Toward the Frontier Fields}

\author{Matthew B. Bayliss\altaffilmark{1,2}, 
Keren Sharon\altaffilmark{3}, and
Traci Johnson\altaffilmark{3}
}

\altaffiltext{1}{Department of Physics, Harvard University, 17 Oxford St., 
Cambridge, MA 02138}
\altaffiltext{2}{Harvard-Smithsonian Center for Astrophysics, 60 Garden St., 
Cambridge, MA 02138}
\altaffiltext{3}{Department of Astronomy, the University of Michigan, 500 Church St. 
Ann Arbor, MI 48109}

\email{mbayliss@cfa.harvard.edu}

\begin{abstract}

We test the effects of varying the cosmological parameter values used in the strong 
lens modeling process for the six Hubble Frontier Fields (HFF) galaxy clusters. The standard 
procedure for generating high fidelity strong lens models includes careful consideration of 
uncertainties in the output models that result from varying model parameters within 
the bounds of available data constraints. It is not, however, common practice to account for 
the effects of cosmological parameter value uncertainties. The convention is to instead 
use a single fiducial ``concordance cosmology'' and generate lens models assuming 
zero uncertainty in cosmological parameter values. We find that the magnification maps 
of the individual HFF clusters vary significantly when lens models are computed using 
different cosmological parameter values taken from recent literature constraints from 
space- and ground-based experiments. Specifically, the magnification maps have 
average variances across the best fit models computed using different cosmologies that 
are comparable in magnitude to -- and as much as $2.5\times$ larger than -- 
the model fitting uncertainties in each best fit model. We also find that estimates of the 
mass profiles of the cluster cores themselves vary only slightly when different 
input cosmological parameters are used. We conclude that cosmological parameter 
uncertainty is a non-negligible source of uncertainty in lens model products for the 
HFF clusters, and that it is important that current and future work which relies on 
precision strong lensing models take care to account for this 
additional source of uncertainty.

\end{abstract}


\section{Introduction}

Strong gravitational lensing enhances the best-available 
observational facilities by naturally zooming in on the distant universe. Massive 
galaxy clusters are the most effective ``natural telescopes'' available to us, 
because they provide high-magnification over relatively large regions of the 
sky ($\sim$1 sq. arcmin). We are entering a new era in which strong lensing 
is transitioning from a niche field into an important piece in the toolkit of observational 
cosmologists. Perhaps the most publicized evidence of this transition is the 
{\it Hubble} Frontier Fields (HFF)\footnote{http://www.stsci.edu/hst/campaigns/frontier-fields/} 
initiative. The HFF are specifically designed to exploit the magnification from strong 
lensing by clusters of galaxies. The HFF, in particular, are intended to probe galaxy populations 
at high redshift that are 10 or more times fainter than the faintest sources detected in 
existing deep field observations, 
and will yield new insights into galaxy evolution studies at high redshift, and the 
properties of galaxies during the epoch of re-ionization. 

Strong lensing holds tremendous potential for enabling studies of the distant universe, 
but using strong lens models introduces new sources of systematic uncertainty 
\citep[e.g.,][]{coe2013,tagore2014,zitrin2014b}. To enable prompt use of new and 
upcoming HFF data to constrain the properties of galaxies in the background 
universe several independent teams were tasked with generating lens models 
using existing archival {\it Hubble} imaging of the HFF 
\citep{bradac2005,bradac2009,liesenborgs2006,jullo2007,jullo2009,merten2009,johnson2014,richard2014}.  
The primary lens model products are the convergence ($\kappa$) and shear ($\gamma$) 
maps, which can be used to construct magnification maps for sources at a given  
redshift, z$_{s}$. These models are a starting point, and are publicly available
\footnote{http://archive.stsci.edu/prepds/frontier/lensmodels/}; 
the first wave of models incorporating new HFF observations are now appearing 
\citep[][]{ishigaki2014,jauzac2014a,jauzac2014b}.

\begin{deluxetable*}{lccc}[h]
\tablecaption{Cosmological Parameter Constraints from the Literature\label{tab:cosmologies}}
\tablewidth{0pt}
\tabletypesize{\tiny}
\tablehead{
\colhead{Source} &
\colhead{$\Omega_{M}$\tablenotemark{a}} &
\colhead{$H_{0}$ {\scriptsize (km s$^{-1}$)} } &
\colhead{Reference} }
\startdata
fiducial ``concordance cosmology'' &  0.300  &  70  &  ---  \\
Planck 2013$+$WP$+$hL$+$BAO  &  0.308 $\pm$ 0.01  &  67.8 $\pm$ 0.8  &  \citet{planck13-XVI}  \\
WMAP-9$+$eCMB$+$H$_{0}$$+$BAO  & 0.286 $\pm$ 0.01 &  69.3 $\pm$ 0.8  &  \citet{hinshaw2013} \\
SPT Clusters$+$WMAP$+$SNe  & 0.255 $\pm$ 0.016   & 71.6 $\pm$ 1.5 &  \citet{reichardt2013}
\enddata
\tablenotetext{a}{~We restrict ourselves to cosmologies that assume a flat geometry, 
so that $\Omega_{\Lambda}=1-\Omega_M$; see the references 
in column 4 for more details.}
\end{deluxetable*}

\begin{figure*}
\centering
\includegraphics[scale=0.715]{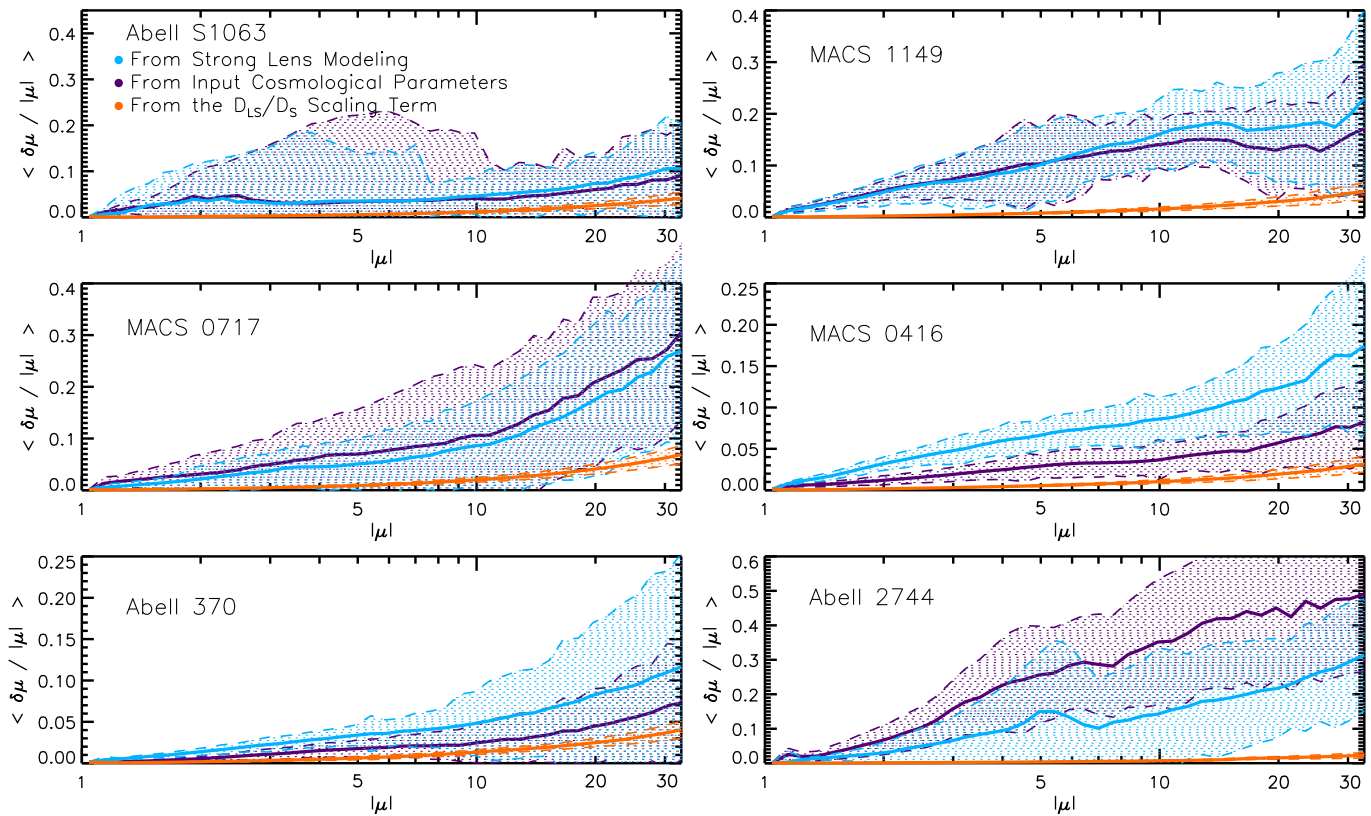}
\caption{\scriptsize{
The fractional magnification uncertainty of individual pixels as a function of the mean 
magnification value in each pixel for each of the six HFF clusters for sources at z $=$ 3. 
The solid lines indicate 
the median fractional uncertainty of all pixels of a given magnification value, and the 
shaded regions indicate the 1$\sigma$ spread. Blue represents the ``statistical'' 
uncertainties from the MCMC minimization over lens model parameters; purple 
represents the uncertainties that result from varying the input 
cosmological parameter values; and orange represents the uncertainties that 
are imposed via the uncertainty in the D$_{LS}$/D$_{S}$ scaling term that is applied 
to the $\kappa$ and $\gamma$ maps to a source-plane at z $=$ 3.}}
\label{fig:magerrmag}
\end{figure*}

The typical approach for generating strong lensing models is to assume a single fiducial 
set of cosmological parameters, for example a flat $\Lambda$ cold dark 
matter ($\Lambda$CDM) cosmology with $H_{0} = 70$ km s$^{-1}$ Mpc$^{-1}$, 
matter density $\Omega_{M} = 0.3$, and $\Omega_{\Lambda} = (1-\Omega_{M}$). 
However, as the precision of the strong lensing models improve they should, at some 
point, become sensitive to the uncertainties in these input cosmological parameter values. 
In this Letter we investigate how strong lensing model uncertainties vary with different 
input cosmological parameter values. We take the input parameter values from three 
recent experiments -- the Wilkinson Microwave Anisotropy Probe 
\citep[WMAP;][]{hinshaw2013}, the South Pole Telescope \citep[SPT;][]{reichardt2013}, 
and the {\it Planck} satellite \citep{planck13-XVI} -- in addition to the fiducial 
``concordance cosmology''. The exact cosmological parameter values 
that we use are summarized in Table~\ref{tab:cosmologies} with more details 
available from the references therein.

\section{Lens Models and Cosmological Parameters}
\label{sec:params}

The cosmological parameters that impact the lens modeling process are those 
that relate directly to computing cosmological distances, i.e., the Hubble constant, 
$H_{0}$ and the matter density $\Omega_{M}$. Here we restrict ourselves to flat 
cosmologies, so that the vacuum energy density, $\Omega_{\Lambda}$, is $1-\Omega_{M}$. 
These parameters influence gravitational lensing via 
the angular diameter distances to the lens, $d_{l}$, the source, $d_{s}$, and between 
the lens and the source, $d_{ls}$ \citep[e.g.,][]{fukugita1992,schneider1992}; different 
values of $H_{0}$ and $\Omega_{M}$ correspond to different geometric distances 
between a gravitational lensing potential and a background source.

\begin{figure*}
\centering
\includegraphics[scale=0.63]{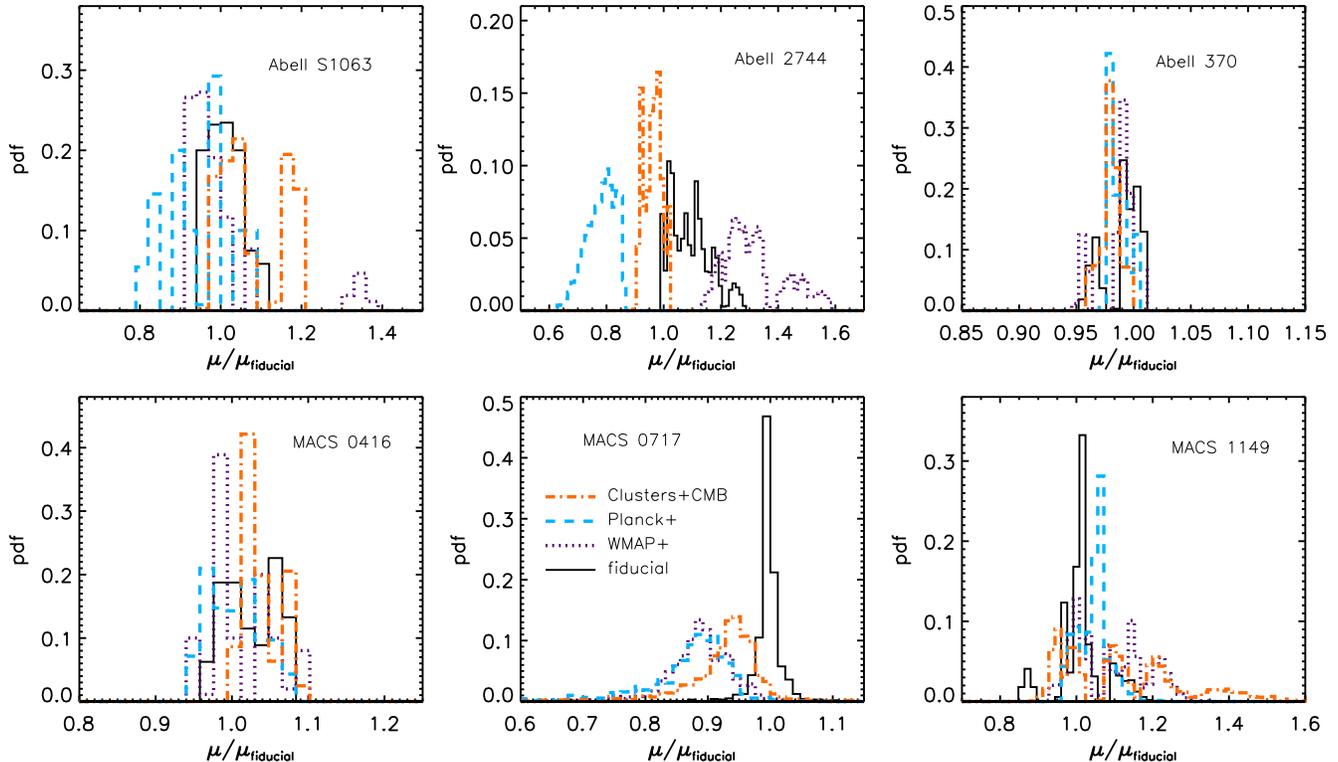}
\caption{\scriptsize{
The distribution of magnification values of pixels in a random position in each of the six HFF 
lens models, separated by input cosmology. At each random position we normalize the 
magnification values of all pixels in a box of size 9$\times$9 pixels (4.5\arcsec~to a side) 
by the magnification value of the central pixel in the best-fit fiducial cosmology. There are clear 
systematic shifts in the distribution of pixel magnification values across a statistical sampling of 
models when different input cosmologies are used. The magnitude and direction of those shifts 
change w/ position across each cluster field; these plots of a random region in each HFF field 
simply serve to illustrate that 
there are systematic shifts in magnification that result from the input cosmology.}}
\label{fig:pixelmagnifications}
\end{figure*}

In the gravitational lens equation,
\begin{equation}
\vec{\beta} = \vec{\theta} - \vec{\alpha}
\end{equation}
the deflection angle, $\vec{\alpha}$, is the difference between the observed and true 
positions on the sky --  $\vec{\theta}$ and $\vec{\beta}$, respectively -- of a background 
source. The deflection angle can be written in terms of the convergence, 
$\kappa (\vec{\theta})$, which is defined as the surface mass density of the lensing 
potential in units of the critical surface mass density, $\Sigma_{crit}$.

\begin{equation}
\div{\vec{\alpha}} = 2 \kappa(\vec{\theta}) ~ 
\end{equation}
where,
\begin{equation}
~ \kappa(\vec{\theta}) = \frac{\Sigma (\vec{\theta})}{\Sigma_{crit}} 
\end{equation}

The critical surface mass density is the surface mass density that is sufficient for a 
gravitational lens at a given redshift, $z_{l}$, to produce multiple images of a 
background source at a given redshift, $z_{s}$:

\begin{equation}
\Sigma_{crit} = \frac{c}{4\pi G}\frac{d_{s}(z_{s})}{d_{l}(z_{l}) d_{ls}(z_{l},z_{s})}.
\end{equation}

Formulated this way one can think of the gravitational deflection angle as a function 
of two terms. The first, $\Sigma (\vec{\theta})$, describes the surface 
mass distribution of the lensing potential, and the second, $\Sigma_{crit}$, 
depends on the source-lens-observer geometry. The second term is sensitive to 
cosmological parameters, as values of $H_{0}$ and $\Omega_{M}$ result in different 
values of the angular diameter distance to the lens, to the source, and between the lens 
and source.

\section{Quantifying the Cosmological Impact on Lens Models}
\label{sec:methods}

Our goal is to quantify the degree to which varying cosmological parameter 
values within their current best-constraints affects measurements of background sources 
that rely on strong lensing models. We narrow our analysis to the HFF because these 
six clusters are among the best studied lenses with an emphasis on their use as precision 
cosmic telescopes. There are two ways in which the input 
cosmological parameters will -- via the distance term in $\Sigma_{crit}$ -- influence 
measurements that rely on lens model products. Firstly, the input cosmology used when 
modeling the strong lensing potential determines basic physical quantities, most notably 
the relationship between angular scale on the sky 
and physical scale in the lens/source planes (e.g., kpc/\arcsec). Secondly, the lensing 
model products are computed for a single source-plane redshift -- for example the 
\citet{johnson2014} lens models use z$_{s} =$9 -- and those products must be scaled 
to the source redshifts of specific individual background galaxies. We investigate 
how each of these effects induce variations in the strong lensing magnification maps. 

We restrict our analysis to the lens 
models of \citet{johnson2014}, which are generated using the \texttt{Lenstool} software 
\citep{jullo2007}; we are explicitly not sampling all sources of uncertainty in the lens modeling 
process. Understanding the ``true'' total lens model uncertainties is the subject of ongoing work 
across the strong lensing community; this Letter is one piece of that larger effort.
We do not account for potential systematic uncertainty due to uncorrelated line-of-sight 
structure \citep{bayliss2014a,daloisio2014}. The magnitude of such line-of-sight effects is not 
well-understood, and including it in the modeling would require significant observational 
follow-up and code development \citep{mccully2014}. 

\subsection{Impact of Cosmological Uncertainty On Magnification Maps}

The first test that we perform is designed to assess the degree to which the input 
cosmological parameters affect strong lensing models. We do this by modeling each 
cluster using four different cosmologies (Table~\ref{tab:cosmologies}) taken from the literature 
to span the range of the current best constraints for $H_{0}$ and $\Omega_{M}$ in a flat 
$\Lambda$CDM cosmology. 
We use identical observational constraints and lens model assumptions as 
\citet{johnson2014}, and generate models for background source redshifts of $z_{s} = 3$. 
For each lens model of each cluster we generate maps that assign magnification values to 
the pixelated sky, with the pixelation of the maps preserved across all models of a given cluster. 

A family of lens models for each cluster in each cosmology is generated from lens model 
parameter values drawn from the MCMC minimization  
\citep[see e.g.,][]{sharon2012,bayliss2014b,johnson2014,sharon2014}. For each cluster in 
each cosmology this provides a statistical range of lens models spanning the 68\% confidence 
region in the MCMC parameter space as traced by $\chi^{2}$. From the family of models 
generated using the fiducial cosmology we measure the 1-$\sigma$ uncertainty in the 
magnification of each pixel as half of the full range spanned across the models that sample 
the 68\% confidence region. We call these uncertainties "statistical uncertainties" throughout 
this letter. We then compute the root mean squared (RMS) scatter in the magnification values 
for each pixel across the best-fit lens models in each of the four input cosmologies, and refer 
to this as the systematic uncertainty that results from varying the input cosmological parameter 
values. 

To quantify the degree to which cosmological parameter uncertainties affect the HFF strong 
lens models we begin by comparing the magnitude of the statistical and systematic 
uncertainties described above. We examine the average fractional uncertainty (statistical 
and systematic) as a function of the magnification for each of the six HFF clusters, and plot 
the results in Figure~\ref{fig:magerrmag}. It is crucial to establish that the scatter in 
magnification across the models with different input cosmologies are not simply the result of 
randomly sampling the statistical uncertainty. To do this we look at the distribution of 
magnification values at random positions on the sky across the full range of statistical 
models generated with each input cosmology for each cluster. Specifically, we select a 
random position and examine the magnification values of all pixels within a 9$\times$9 
pixel box (4.5\arcsec$\times$4.5\arcsec) centered on that position, and plot the distribution 
of all pixel magnification values in that region, across all statistical models in each of the four 
cosmologies. 

In Figure~\ref{fig:pixelmagnifications} we show the results for a single random location in 
each HFF cluster field; the positions used here are just one randomly selected realization 
of $\sim$100 such draws. The probability distribution of magnification values does change 
with the input cosmology -- sometimes very dramatically and sometimes only weakly. The 
fact that the distribution of magnification values at a given position shifts {\it systematically} 
with input cosmology confirms that uncertainty in cosmological parameter 
values maps directly into a systematic uncertainty in lens models. 

\begin{figure}
\centering
\includegraphics[scale=0.57]{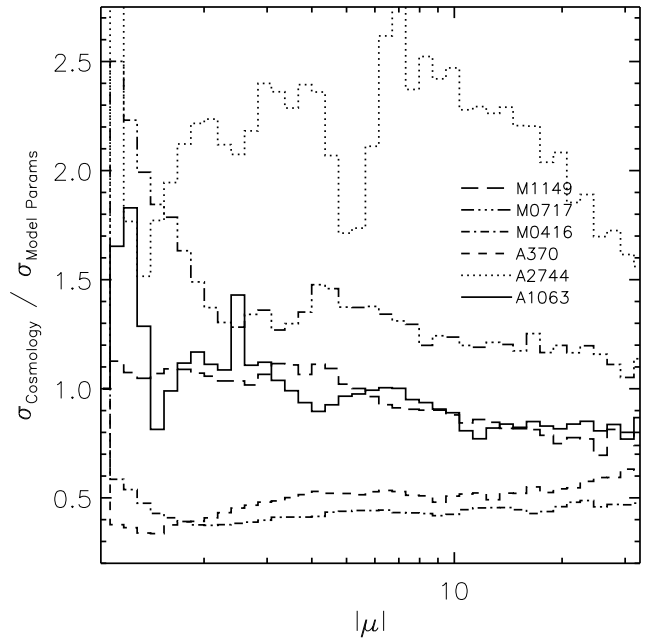}
\caption{\scriptsize{
Ratio of median magnification uncertainties as a function of magnification that result 
from lens models with different input cosmologies vs. the statistical uncertainties from 
the strong lens models; each of the six HFF is plotted 
individually.}}
\label{fig:errorratio}
\end{figure}

The relative scale of cosmological noise in the magnification 
maps for the HFF clusters are shown in Figure~\ref{fig:errorratio}, where we plot the 
ratio of the fractional magnification uncertainties for each of the two cases -- this is simply 
the ratio of the two solid lines plotted in each panel of Figure~\ref{fig:magerrmag}. 

\subsection{Impact of Cosmological Uncertainty When Scaling Lens Models to Arbitrary Source Redshifts}
\label{sec:scalecosmo}

In addition to the modeling process, we must also assess the degree to which cosmological 
parameter uncertainties affect the ability to precisely scale lens model outputs to background 
sources at arbitrary redshifts. The magnification is computed directly from the maps for 
$\kappa$ and $\gamma$ \citep{schneider1992}: 
\begin{equation}
\mu = \frac{1}{(1 - \kappa)^{2} - \gamma^{2}}
\end{equation}
where both $\kappa$ and $\gamma$ scale proportionally to the distance term, 
$d_{ls}/d_{s}$. Here we assess the uncertainty in the distance ratio, $d_{ls}/d_{s}$, due to 
cosmological parameter uncertainty. Uncertainty in this distance ratio adds additional 
noise into magnification maps that are scaled to an arbitrary source redshift. We do this 
by calculating the fractional variation in the value of $d_{ls}/d_{s}$ using the 
fiducial ``concordance'' parameter values as the control distance ratio -- $d_{ls,0}/d_{s,0}$ -- 
and compare this fiducial value to the distance ratio evaluated for other cosmological 
parameter values drawn from constraints in the literature (Table~\ref{tab:cosmologies}). 
This fractional uncertainty is computed as, 
\begin{equation}
\frac{d_{ls}(z_{s})/d_{s}(z_{s})}{d_{ls,0}(z_{s})/d_{s,0}(z_{s})}.
\end{equation}
We then apply the uncertainty in this distance ratio to the Sharon v2 $\kappa$ and $\gamma$ 
maps available from the HFF website to generate a range of scaled maps; from these maps we 
compute the RMS uncertainty in the magnification values for each pixel across these scaled maps 
and include the resulting fractional magnification uncertainty vs magnification in 
Figure~\ref{fig:magerrmag}. This uncertainty is sub-dominant to the other statistical and systematic 
uncertainties explored above, primarily because factors of $H_{0}$ 
cancel out when calculating the distance ratio, so that only the matter 
density, $\Omega_{M}$ matters, and because both of the relevant angular diameter distances, 
$d_{ls}$ and $d_{s}$, change in the same way with changes in $\Omega_{M}$.

\subsection{Impact of Cosmological Uncertainty On Lensing Reconstructed Mass Maps}

\begin{figure}
\centering
\includegraphics[scale=0.49]{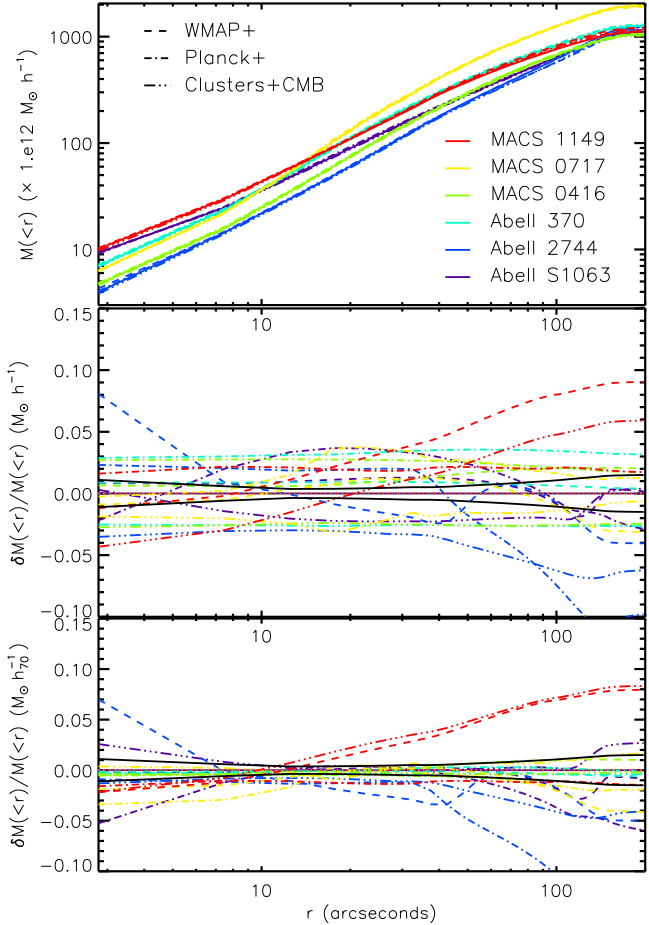}
\caption{\scriptsize{
Enclosed mass as a function of radius and enclosed mass residuals 
(relative to the fiducial cosmology, see Table~\ref{tab:cosmologies}) for each of the six 
HFF clusters, plotted for the best fit mass maps in each of the four different 
cosmologies used here. The bottom two panels show the residuals in physical 
units (middle panel in units of M$_{\odot}$ h$^{-1}$) and in common units normalized to 
the fiducial H$_{0}$ value (bottom panel in M$_{\odot}$ h$^{-1}_{70}$). The residuals 
in the bottom panel most accurately reflect the cosmological uncertainties in the enclosed 
mass that result from lens model variations with cosmology. The statistical uncertainties 
in the enclosed mass are over-plotted as the solid black lines in the two residual panels, 
and are small ($<$2\%) at all radii; the cosmological uncertainties are everywhere as 
large or larger than the statistical uncertainties.}}
\label{fig:mass}
\end{figure}

In this section we quantify the effect of varying the 
input cosmological parameters on the resulting mass maps for the lensing clusters. We 
take the output mass maps for the best fit lens model in each of the four cosmologies 
(Table~\ref{tab:cosmologies}) and measure the total enclosed mass vs. projected radius for 
each of the six HFF clusters, along with the residuals relative to the fiducial 
cosmology; the results are shown in Figure~\ref{fig:mass}. We also estimate the statistical 
uncertainty in the mass profiles of the HFF clusters using the same lens models 
analyzed in previous sections. The 1-$\sigma$ statistical uncertainty here is simply the half 
range of the value of the mass profile as computed from the full range of statistical lens 
models drawn from the 68\% confidence region of the MCMC chains. These statistical 
uncertainties are plotted in the middle and lower panels of Figure~\ref{fig:mass} along with 
the cosmological residuals. 

Residuals in the enclosed mass (in physical units) are typically $\pm5$\% and remain fairly 
constant over two decades in projected radius. However, some component of these residuals 
are simply the result of the differences in the factor of 
h$^{-1}$ that is multiplied into cosmological mass measurements. When we factor out this 
h$^{-1}$ from the masses from different cosmologies the mass residuals shift to become 
smaller ($\sim$1-2\%) at the radii where the best strong lensing constraints lie 
(2\arcsec$\lesssim$r$\lesssim$20\arcsec), with a few clusters having larger residuals at 
other radii.

One of the great strengths of strong lensing mass measurements is its precision, with 
typical statistical uncertainties of $\sim$1\% (Figure~\ref{fig:mass}). Systematic cosmological 
mass residuals are as larger or larger ($\sim$1-2\%) at radii where strong lensing constraints 
are typically used, but both statistical and cosmological uncertainties are very small in this 
region. While cluster mass measurements are not the primary science driver of 
the HFF, this remains one of the more powerful applications of strong lensing clusters 
\citep{comerford2007,oguri2009a,oguri2012,gralla2011,merten2014}, and it would be 
valuable for future analyses to consider the systematic uncertainty on cluster masses 
that are derived from strong lensing models.

\section{Discussion and Implications}
\label{sec:discussion}

Varying the input cosmological parameters for the strong lens modeling process results 
in significant magnification uncertainties for all six of the HFF clusters. From 
Figures~\ref{fig:magerrmag}~and~\ref{fig:errorratio}~we see that some of the clusters 
seem to be more strongly affected by cosmology than others. For example, Abell S1063, 
Abell 2744, MACS J0717, and MACS J1149 all have cosmological magnification 
uncertainties that are generally equal to or greater than the statistical uncertainties in the 
lens models, while Abell 370 and MACS J0416 have average cosmological magnification 
uncertainties that are only $\sim$40-60\% those the lens model parameter uncertainties. 
There are several possible drivers for increased uncertainty from cluster to cluster 
(e.g., number of constraints, the positional distribution of constraints, and availability of  
spectroscopic redshifts; Johnson et al.~in prep), but we do note that clusters with a high 
fraction of arcs with spectroscopic redshifts have lower cosmological uncertainties 
compared to the statistical uncertainties. Specifically, 56\% of the background sources 
for Abell 370 and 66\% for MACS J0416 have spec-z's, whereas the background sources 
for the other four clusters all have between just 25-38\% spec-z's \citep[Table 3 in][]{johnson2014}.

The broad results of our analysis here does strongly argue that it is not appropriate to assume  
that lens model parameter uncertainties dominate the error budget of precision strong lens 
models. There are already a number of early results using the first-pass HFF lens models to 
measure the intrinsic properties of distant background galaxies 
\citep{atek2014a,atek2014b,bradley2014,ishigaki2014,kawamata2014,mcleod2014,monna2014,schmidt2014}. 
Cosmological parameter uncertainty will fundamentally contribute at some level to the 
systematic uncertainty in strong lens modeling, and our work here demonstrates that the 
cosmological contribution is likely at a level that cannot be ignored.
Precisely quantifying these effects is certain to be sensitive to the modeling methodology; 
our results are specific to the lens models published in \citet{johnson2014}, and it 
would be up to individual lens model teams to fold in methods that allow for a range of input 
cosmologies. A true characterization of the magnification uncertainty marginalized 
over both cosmology and lens model parameters will require the investigation of 
cosmological parameter constraints in the MCMC minimization code.  Looking ahead it 
is important that the strong 
lensing community continue the trend toward producing public lens models with 
comprehensive assessments of all relevant systematics, which certainly includes adopting 
methodologies that marginalize lens model uncertainties over a range of cosmological 
parameter values that span the current best-constraints. 

One additional, somewhat tangential, implication of these results is that we are 
entering an era in which precision strong lensing maps of the HFF clusters might be used to 
provide independent constraints on geometric cosmological parameters. \citet{jullo2010} first 
used the deflection constraints from multiply imaged sources in precision lens models to 
constrain cosmological parameter values, but it is not clear that the HFF lens models will 
necessarily be able to do better than previous work by \citet{jullo2010}. 

\section{Conclusions}
\label{sec:conclusions}

Our results indicate that cosmological parameter uncertainties 
{\it do} contribute to the noise in magnification maps recovered for the HFF clusters at 
levels that are often comparable in magnitude to the statistical uncertainties in the lens 
models, and that they also impact strong lensing cluster masses at a similar level to the 
statistical modeling uncertainties. The prospect of producing competitive constraints on 
cosmological parameters from the deflection of light via strong lensing also is becoming 
interesting as the number of lenses with numerous constraints increases. In the new era 
of precision strong lens modeling, it is important that {\it all} source of systematic 
uncertainty be considered when totaling up the error budgets of strong lensing models 
for systems that are intended to be used as precision gravitational lenses -- such as the HFF. 

\acknowledgments{
This work utilizes gravitational lensing models that were generated as a part of the 
HST Frontier Fields program conducted by STScI. STScI is operated by the Association 
of Universities for Research in Astronomy, Inc. under NASA contract NAS 5-26555. The 
lens models are hosted on the Mikulski Archive for Space Telescopes (MAST).
The authors thank the referee, Dan Coe, for his thoughtful and helpful feedback. MBB 
acknowledges support from the NSF through grant AST-1009012
and from NASA through grant HST-GO-13639.01.}


\end{document}